\documentclass[aps,pra,twocolumn,showpacs]{revtex4}
\usepackage{graphicx,amsmath,amssymb,color}
\usepackage[normalem]{ulem}
\usepackage{amsfonts}
\usepackage[toc,page]{appendix}
\usepackage[ocgcolorlinks,colorlinks=true,linkcolor=blue,citecolor=red]{hyperref}
\usepackage{latexsym}
\usepackage{amsfonts}
\usepackage{algpseudocode}
\usepackage{amsthm}
\usepackage{mathrsfs}
\usepackage{natbib}
\usepackage{color,verbatim}
\usepackage{psfrag}

\newcommand{\rc}{\mathbf{\rho}}

\def\beq{\begin{eqnarray}}
\def\eeq{\end{eqnarray}}

\begin{document}
\date{\today}
\addtolength{\topmargin}{0.75in}

\bibliographystyle{unsrt}
\title{A Direct Mapping of Max k-SAT and High Order Parity Checks to a Chimera Graph}

\author{N.\ Chancellor$^{\dagger,1}$, S.\ Zohren$^{\dagger,2,3}$, P. A.\ Warburton$^{4,5}$, S.\ C.\ Benjamin$^{2}$, S. Roberts$^{3}$}

\affiliation{$^1$Department of Physics, Durham University, South Road, Durham, UK   \\
$^2$Department of Materials, University of Oxford, Parks Road, Oxford, UK \\
$^3$Department of Engineering Science, University of Oxford, Parks Road, Oxford, UK \\
$^4$London Centre for Nanotechnology 19 Gordon St, London, UK  \\
$^5$Department of Electronic and Electrical Engineering, UCL, Torrington Place, London, UK \\
$^\dagger$Authors contributed equally. Correspondance to: nicholas.chancellor@gmail.com
}

\begin{abstract}
We demonstrate a direct mapping of max $k$-SAT problems (and weighted max $k$-SAT) to a Chimera graph, which is the non-planar hardware graph of the devices built by D-Wave Systems Inc. We further show that this mapping can be used to map a similar class of maximum satisfiability problems where the clauses are replaced by parity checks over potentially large numbers of bits. The latter is of specific interest for applications in decoding for communication. We discuss an example in which the decoding of a turbo code, which has been demonstrated to perform near the Shannon limit, can be mapped to a Chimera graph. The weighted max $k$-SAT problem is the most general class of satisfiability problems, so our result effectively demonstrates how any satisfiability problem may be directly mapped to a Chimera graph. Our methods faithfully reproduce the low energy spectrum of the target problems, so therefore may also be used for maximum entropy inference.

\end{abstract}

\maketitle

\section*{Introduction}

\begin{figure}
\begin{centering}
\includegraphics[width=7cm]{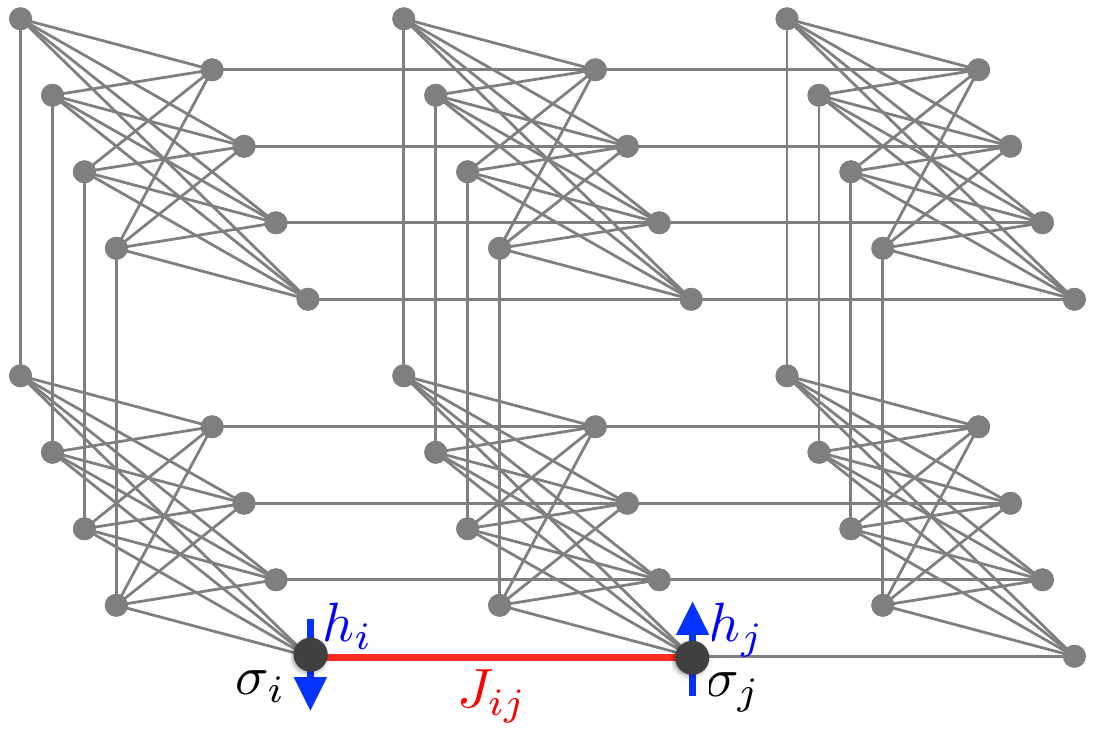}
\par\end{centering}
\caption{A region  of a Chimera graph containing $3\times 2$ unit cells. Each vertex corresponds to a spin variable. One can adjust the magnetic fields $h_i$ at each spin as well as the pair-wise couplings $J_{ij}$ between spins which are adjacent to each other in the graph.  \label{fig-empty-chimera} }
\end{figure}

Many interesting computer science problems have been shown to be directly mappable to finding the ground state of a Ising spin model on the hardware graph of the devices by D-Wave systems Inc. \cite{D-Wave}, the Chimera graph (see Figure \ref{fig-empty-chimera}). Examples include Maximum-Weight Independent Set, Exact Cover, and 3-SAT Problems \cite{Choi2004}. Technically, by virtue of being NP-complete, an efficient mapping of any one of these problems implies that any other NP-complete problem can also be mapped to finding the ground state of a spin model on the Chimera graph. In practice however, such an indirect mapping is likely to be  impractical given the current limitations of size and energy scales on real devices. It is for this reason that we are interested in direct mappings of interesting problems onto the Chimera graph, and why a direct mapping of a very general problem such as max $k$-SAT is of interest.

The ultimate reason why we are interested in mapping to the Chimera graph is to more efficiently map the 2-body Ising models which implement these problems to the hardware graph of the D-Wave annealers, although modifications to the ideas given here may be useful for mapping to other graphs which are made of tilings of locally non-planar graphs.  Much of the attention in developing minor embeddings for the Chimera graph has been on increasing connectivity. In particular, a fully connected graph can always be mapped to a Chimera graph using minor embedding \cite{Choi2011}. Recently, this focus has also included purpose-built architectures different from the Chimera graph \cite{Lechner2015} (see also \cite{Albash2016} and \cite{Andrea} for related work). Here our focus is on higher order terms which are necessary to implement clauses and parity terms which involve more than two variables. However, as we will see later we will be using some of the minor embedding techniques for fully connected graphs to construct the embeddings of such terms.

There have been many promising advances in quantum annealing, since the idea that quantum fluctuations could help explore rough energy landscapes \cite{Ray1989}, through the algorithm first being explicitly proposed \cite{Finilla1994}, further refined \cite{Kadowaki1998},  and the basic concepts demonstrated experimentally in a condensed matter system \cite{Brooke1999}. Recently both entanglement \cite{Lanting2014} and tunneling \cite{ Boixo2016,Boixo2014} have been experimentally demonstrated on programmable annealing processors. Given these encouraging results, it is desirable to propose new problem types for these machines to solve. This is interesting both from the viewpoint of possible eventual commercial applications, as well as providing access to new problems sets for benchmarking. For an overview of some aspects of quantum annealing, please see \cite{Das2008}.

Another method of mapping problems onto the Chimera graph, is the one employed by \cite{Bian2014}. This method uses numerical algorithms, often heuristic ones in practice, to  map the problems. For reasons which we will discuss later, the low density parity check code (LDPC) decoding done in \cite{Bian2014} can be thought of as a mapping of a weighted SAT problem into the Chimera graph. One crucial difference however, is that this LDPC decoding \emph{cannot} be thought of as mapping a max-SAT problem, as our method does. For other examples of problem mappings see \cite{Biamonte2008,Whitfield2012}.

Also unlike this mapping, our technique can not only be used for optimization tasks but also for sampling.  This is important considering that recently there has been much interest in using D-wave for sampling applications, especially in the context of training Boltzmann machines \cite{Boltzmann1,Boltzmann2,Boltzmann3}, but also in the context of message decoding \cite{Chancellor2016}.  There are also many other examples in which maximum entropy inference, which relies on sampling approximate thermal distributions can be applied in fields as varied as finance \cite{Mistrulli2011},  ecology \cite{Phillips2006}, and computational linguistics \cite{Berger1996}. Even more powerful problem embeddings can probably be created by combining the ideas presented here with the powerful numerical techniques used in \cite{Bian2014} .

We should also compare our paper to other recent work on using quantum annealing to solve satisfiability problems \cite{Azinovic2016}. This work examines the use of quantum annealers to build SAT filters, which require a relatively large number of disparate solutions of a SAT problem to construct. This work finds that quantum annealing is not a suitable method compared to classical methods. Building a SAT solver is significantly different than solving such a problem directly, as we discuss in this paper, and is most appropriately classified as a variant of \#SAT. Furthermore, SAT filters are not directly applicable to max-SAT problems as we study here, and so should be regarded as related, but very much distinct. It is worth remarking that the methods we give here could still potentially be useful for \#SAT type problems such as SAT filter construction if the annealer were used to perform a hybrid algorithm as suggested in \cite{Neven2016,Chancellor2016a,Chancellor2016b} rather than the standard quantum annealing algorithm.

For a review on boolean satisfiability, we point the reader to \cite{Malik2009}, and for max-SAT in particular to \cite{Stutzle2001}.


\section*{Implementing Clauses}

\subsection*{Basic Operations} \label{sub-simple}

Any Boolean clause can always be written out as logical AND operations performed on strings of logical OR operators performed on bit values or the logical negation of bit values, e.g.\ $(a_1\, \mathrm{OR} \, a_2 ...)\mathrm{AND} (\mathrm{NOT}\, a_1\, \mathrm{OR} \, a_5 ... )$. In the following we denote $\mathrm{AND}$ by $\wedge$, $\mathrm{OR}$ by $\vee$ and negation by $\neg$. A general clause is thus of the form
\begin{equation}
(a^{(1)}_1 \vee a^{(1)}_2... )\wedge (a^{(2)}_1 \vee a^{(2)}_2... ) \wedge ...,
\end{equation}
where $a^{(l)}_{i}\in \{a_1, a_2, a_2,...\} \cup \{\neg a_1, \neg a_2,\neg a_3,...\}$. All that is needed to implement arbitrary Boolean clauses is therefore the ability to implement clauses of the form $a^{(l)} := (a_1^{(l)} \vee a_2^{(l)}... )$. To implement a SAT problem in terms of energy computation, we could construct such a term by enforcing a penalty of the form,

\begin{equation}
Pen(\{a^{(l)}\})\begin{cases}
\geq g  & a^{(l)}_i=0 , \, \, \forall i \\
=0 & \textrm{otherwise.}
\end{cases}\label{eq:penalty}
\end{equation}

One can then construct a SAT problem by summing many such penalties and obtaining an energy $E=\sum_l Pen(\{a^{(l)}\})$. If one or more bit-strings exist where $E=0$ then a set of clauses is satisfiable, but otherwise it is not. In the case where the clauses are satisfiable, the bit-strings which yield $E=0$ are the ones which satisfy the clauses. However, because the penalties are unknown and are set to simply be an arbitrary value greater than or equal to $g$, the energies of states with $E>0$ are meaningless. If no bit-string can satisfy all clauses, the lowest $E$ state is \emph{not} necessarily the one which satisfies the most clauses, and this is therefore \emph{not} a valid construction of a max-SAT problem. 
 
 However, if we consider terms which give all violated clauses the same energy penalty,
 
\begin{equation}
Spec(\{a^{(l)}\})=\begin{cases}
g & a^{(l)}_i=0 \, \, \forall i \\
0 &  \textrm{otherwise},
\end{cases}.\label{eq:spectrum}
\end{equation} 
and similarly construct a total energy $E=\sum_l Spec(\{a^{(l)}\})$, then for $g>0$ the minimum energy bit-string will always be the one which satisfies the most clauses, regardless of whether all clauses can be simultaneously satisfied. An energy penalty of this form therefore is a valid expression of a max-SAT problem. In this letter we show not only a natural way to express such penalties in terms of the Ising model, but also how such terms may be efficiently embedded into the D-Wave Chimera graph. It is worth pointing out that this can easily be even further generalized to a weighted version of the max $k$-SAT problem if a different value of $g$ is chosen for different clauses $a^{(l)}$ in a controlled rather than arbitrary way.

To move from logical values to spin variables, we map each logical variable $a_i=0$ to a spin variable with value $\sigma_i^{z}=-1$ and each logical variable $a_i=1$ to a spin variable with value $\sigma_i^{z}=+1$. Negation of the logical variables is then implemented through gauges on the spin variables. More precisely, we map $a_i$ to $c(i)\sigma_i^{z}$ with $c(i)=1$ and $\neg a_i$ to $c(i)\sigma_i^{z}$ with $c(i)=-1$. Formally, we thus identify $a_i =\frac{1}{2} (1+ c(i)\sigma^z_i)$ with $c(i)=1$ and $\neg a_i = \frac{1}{2} (1+ c(i)\sigma^z_i)$ with $c(i)=-1$. 

\begin{figure}
\begin{centering}
\includegraphics[width=6cm]{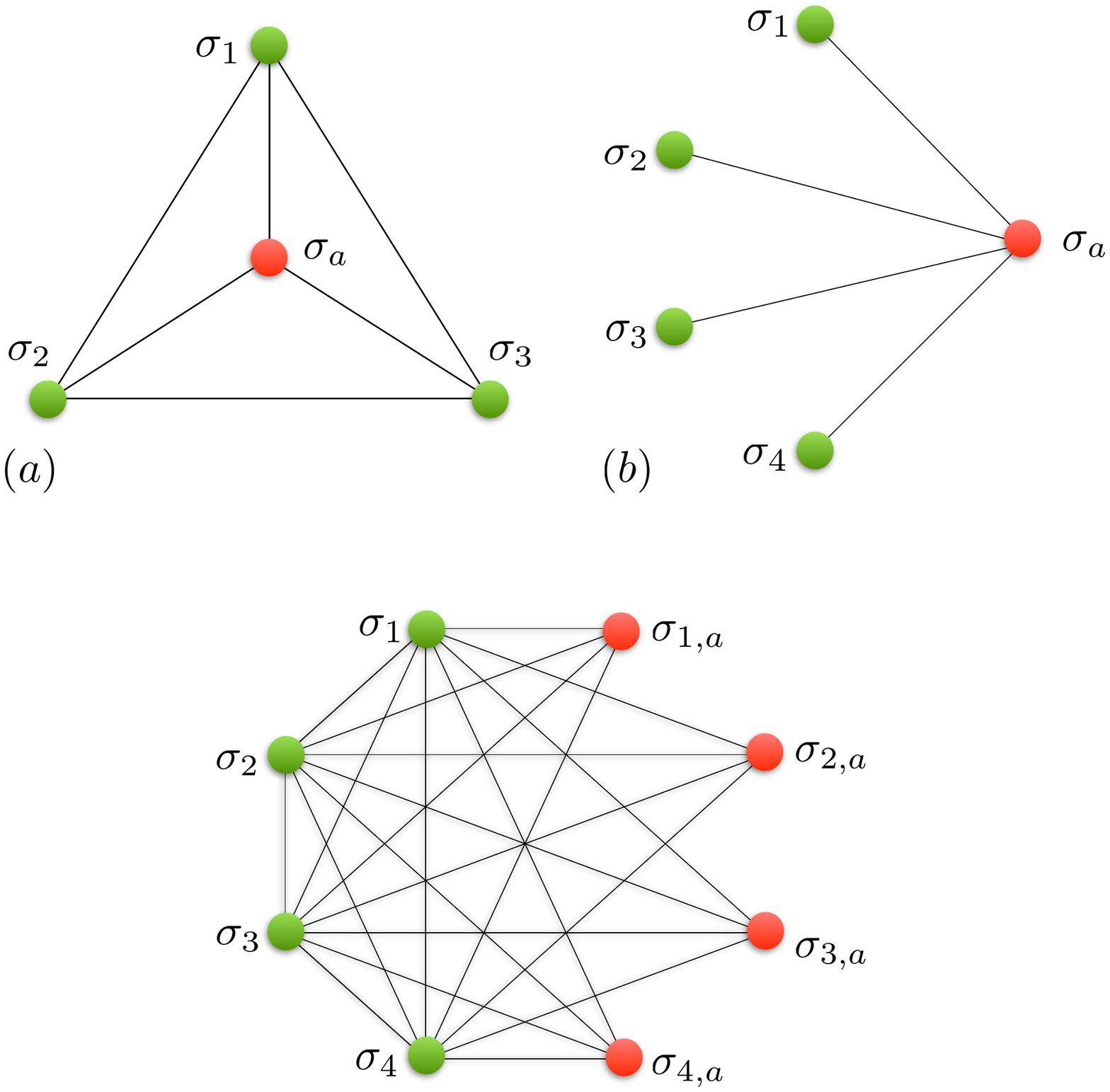}
\par\end{centering}
\caption{Illustration of the spin graph corresponding to the implementation of the clause $\sigma_1 \vee \sigma_2 \vee \sigma_3 \vee \sigma_4 $, as well as the parity checking clause $\sigma_1 \oplus \sigma_2 \oplus \sigma_3 \oplus \sigma_4 $. Logical spin variables are shown as green vertices, ancillae as red vertices, non-zero couplings are shown as black edges and magnetic fields are not shown. Both clauses for OR and XOR only differ in the value of their fields.\label{fig-schematic} }
\end{figure}

 We now consider how to implement a single SAT clause using an Ising spin Hamiltonian which might be implemented on an annealing machine like those of D-Wave systems Inc. Consider a Hamiltonian for spin variables of the form


 \begin{eqnarray}
\mathcal{H}_{\mathrm{clause}}^{(2)}&=& J \sum_{i=1}^k \sum_{j =1}^{i-1}c(i) c(j)  \sigma^{z}_i \sigma^{z}_j + h \sum_{i=1}^k  c(i)\,\sigma^{z}_i + \nonumber \\ 
&& \quad + J^a \sum_{i=1}^k \sum_{j=1}^k  c(i)  \sigma^{z}_i \sigma^{z}_{j,a} +   \sum_{i=1}^k  h_i^a \sigma^{z}_{i,a}.
\label{H2localemb_coup}
\end{eqnarray}
 in which up to the gauge choice $c(i) \in \{ -1,1\}$, $k$ \emph{logical} spin variables $\sigma^{z}_i$ are coupled to $k$ ancilla spin variables -- ancillae for short. The Hamiltonian above is similar to the Hamiltonian presented in \cite{patent_paper} and in fact both are special cases of a more general construction presented below. To implement a single clause, we set  $J=J^a$, $h=-J^a$, as well as $h_{i}^a=-J^a(2i-k)+q_i$ with 
 \begin{eqnarray}
q_i &=& 
\begin{cases}
  g/2    & i=1, \\
 0   & \text{otherwise},
\end{cases} \label{choices}
\end{eqnarray}
where $g/2\ll J_a$. By the symmetry of this Hamiltonian, the effective energy penalty on the ancillae from the logical spin variables for being up or down will depend only on the total number of logical spin variables which are in agreement with the gauges, and not the specific arrangement.  With this choice of $q_i$, all bit-strings will have the same energy unless $c(i)\sigma^z_i=-1, \forall i$ in which case the energy will be greater by $g$. Thus, up to an irrelevant total energy shift, \eqref{H2localemb_coup} generates a single penalty term of the form of \eqref{eq:spectrum}. Table \ref{table} illustrates an example for four bits with all gauge values set to 1. Figure \ref{fig-schematic}  shows the connectivity of the corresponding abstract spin graph. Shown are the four logical spin variables in green and the ancillae in red.

\begin{table}[t]
\begin{center}
\begin{tabular}{| l | c |  c |}
\hline
Logical bit values & Ancilla values  & E  \\
\hline 
$1111 $ & $0000 $ & $0$  \\
$0111$, $ 1011$, $1101$, $1110$
  & $0001$ & $0$  \\
$0011$, $0101$, $0110$, $1001$,
$1010$, $1100$
  &  $0011$  & $0$ \\ 
$1000$, $0100$, $0010$, $0001$ 
   & $0111$ & $0$  \\
$0000$  & $1111$ & $g$ \\
\hline
\end{tabular}
\end{center}
\caption{Summary of the possible configuration of bit-strings for the OR clause $a_1 \vee a_2 \vee a_3 \vee a_4 $ together with the corresponding ancilla value and the energy (up to a constant offset).\label{table}}
\end{table}%

We have demonstrated above how to implement a single clause of the form $a^{(l)} = (a_1^{(l)} \vee a_2^{(l)}... )$. Using this construction we can now implement clauses of the form $(a^{(1)}_1 \vee a^{(1)}_2... )\wedge (a^{(2)}_1 \vee a^{(2)}_2... ) \wedge ...$ by superimposing the construction of the individual clauses $a^{(l)}$ on a common set of logical spin variables. To do so, we include the gauge variables in the coupling constants, i.e.\ associated with the clause $a^{(l)}$ the fields are $h_i^{(l)} = c^{(l)}(i)  h^{(l)}$ and the couplings are $J_{ij}^{(l)} = c^{(l)}(i) c^{(l)}(j) J^{(l)}$. The total fields applied to a spin variable $\sigma^z_i$ are then the sum of the field contribution from each clause, i.e.\ $h_{i} = \sum_l h_{i}^{(l)}$ and similarly for the couplings $J_{ij} =\sum_l J_{ij}^{(l)}$. Note that the ancillae cannot be superimposed.

\subsection*{XOR Clauses and Parity Checks} \label{sub-parity_check}

 We now describe another construction where instead of OR we have clauses constructed out of XOR relations. Such clauses are important in many message decoding applications. Furthermore, they provide an alternative method for implementing the above clauses. In particular, instead of implementing clauses of the form $a^{(j)} := (a_1^{(j)} \vee a_2^{(j)}... )$ and superimposing those to construct the bigger problem $(a^{(1)}_1 \vee a^{(1)}_2... )\wedge (a^{(2)}_1 \vee a^{(2)}_2... ) \wedge ...$, one can also implement bigger clauses directly. To do so, note that a logical AND operation can  be expressed as a product of two such operators $a_i\land a_j=a_i\, a_j$ and the OR operation can be written as $a_i\lor a_j = a_i + a_j-2\, a_i\,a_j$. Inserting the mapping to spin variables, $a_i \rightarrow \frac{1}{2}(\sigma^z_i+1)$,  we see that any Boolean clause can be rewritten  in terms of spin variables, and the penalty Hamiltonians can alternatively be constructed using the methods given in \cite{patent_paper} in terms of multi-body terms. A product of spins corresponds to a  \emph{parity term} or \emph{parity checking clause}, which is represented by an  XOR relation in the Boolean language, i.e. a  term of the form $(a_1^{(l)} \oplus a_2^{(l)} \oplus a_3^{(l)} ... )$, where $\oplus$ denotes bitwise addition. Thus we have the mapping,
  \begin{eqnarray}
 a_1^{(l)} \oplus a_2^{(l)} \oplus ...\oplus a_k^{(l)}  \quad \leftrightarrow \quad c(1) \sigma^z_1c(2)  \sigma^z_2 ... c(k) \sigma^z_k 
 \end{eqnarray}
 The spectrum $Spec(\{ a_1^{(l)} \oplus ...\oplus a_k^{(l)} \})$ is mapped to the spectrum of the Hamiltonian
\begin{equation}
\mathcal{H}_k =\frac{g}{2} c(1) \sigma^z_1 \, \, ... \,\, c(k) \sigma^z_k.
\end{equation}

The abstract connectivity graph of this Hamiltonian is shown in Figure \ref{fig-schematic} for $k=4$. 
We can use the Hamiltonian \eqref{H2localemb_coup} to reproduce the (low-energy) spectrum of $\mathcal{H}_k $ where, as before $J=J^a$, $h=q_0-J^a$, as well as $h_{i}^a=-J^a(2i-N)+q_i$, but $q_i$ is instead chosen as
 \begin{eqnarray}
q_i &=& 
\begin{cases}
 q_0 +  g/2    & \text{$N-i$ is odd}, \\
 q_0 -  g/2   & \text{$N-i$ is even}.
\end{cases} \label{choices_parity}
\end{eqnarray}
with $g/2<q_0\ll J_a$. This assignment of coupling constants is the same as used in \cite{patent_paper}.

It is also worth briefly pointing out that the mapping between logical and spin variables discussed above is completely invertible. 
Any problems which can be expressed as a sum over products of spin operators can therefore be written as a  weighted sum over Boolean clauses plus a constant. Problems expressed as sums of parity checking clauses  which are each weighted equally can therefore be regarded as a version of max $k$-SAT. In an upcoming work \cite{in_prep} we analyze the potential of using this way of implementing parity checking terms in terms of these weighted max-SAT implementations for inference in message decoding problems, in particular when applied to Low Density Parity Check (LDPC) codes and turbo codes. The advantage of our construction in comparison to an earlier construction of LDPC codes \cite{Bian2014} is that it not only reproduces correctly the ground state but also the low-energy spectrum, thus permitting us to do sampling applications, such as the maximum entropy inference discussed in \cite{Chancellor2016}. While we leave the extensive analysis of various message decoding problems to the upcoming work \cite{in_prep}, later in this manuscript  we give a brief description of how to implement turbo codes.

\subsection*{Other Clauses} \label{more_complex}

We have already demonstrated that a simple clause can be expressed as an energy penalty by using $k$ ancillae. This therefore allows a natural construction of any complex $k$ bit clause by simply examining every possible bit string and penalizing it if it violates the clause. For large $k$ this method is rather inefficient for arbitrary  clauses, however, as the number of bit-strings which must be examined potentially grows as $k2^k$. For example, expressing a parity checking clauses using individual SAT clauses would require $k\,2^{\lfloor k/2 \rfloor}$ ancillae. 

It is therefore worthwhile to briefly address how one might go about constructing methods for implementing clauses which cannot be easily expressed using the methods previously discussed more efficiently. Firstly we note that the previous constructions can be generalized by choosing $q_i$ in \eqref{choices} and \eqref{choices_parity} differently. By doing this we can implement any clause which is symmetric under permutation of any of the bits. Such a clause will be defined by a vector $f_i\in \{0,g/2\}$ with $i=0,...,k$ which is $0$ if the set of bit-strings with $i$ bits equal to $1$ satisfies the clause and $g/2$ otherwise. We now define
 \begin{eqnarray}
q_i &=& q_0+ f_{i-1}-f_{i}, \quad i=1,...,k \label{choices_symmetric}
\end{eqnarray}
with $g/2<q_0\ll J^a$. We further observe that the gauges $c(i)\in \{-1,1\}$ in \eqref{H2localemb_coup} allow us to define such clauses which are symmetric \emph{in any gauge}, and potentially to combine more than one of this type of clause constructed in multiple different gauges.
These more complicated constructions should allow many different clauses involving relatively large numbers of bits to be implemented more efficiently than the method given earlier. In practice one would probably want to construct numerical algorithms to find more optimal implementations of arbitrary high $k$ clauses, but this is beyond the scope of the current letter.

\subsection*{Special Cases} \label{sub-special}

\begin{figure}
\begin{centering}
\includegraphics[width=8.5cm]{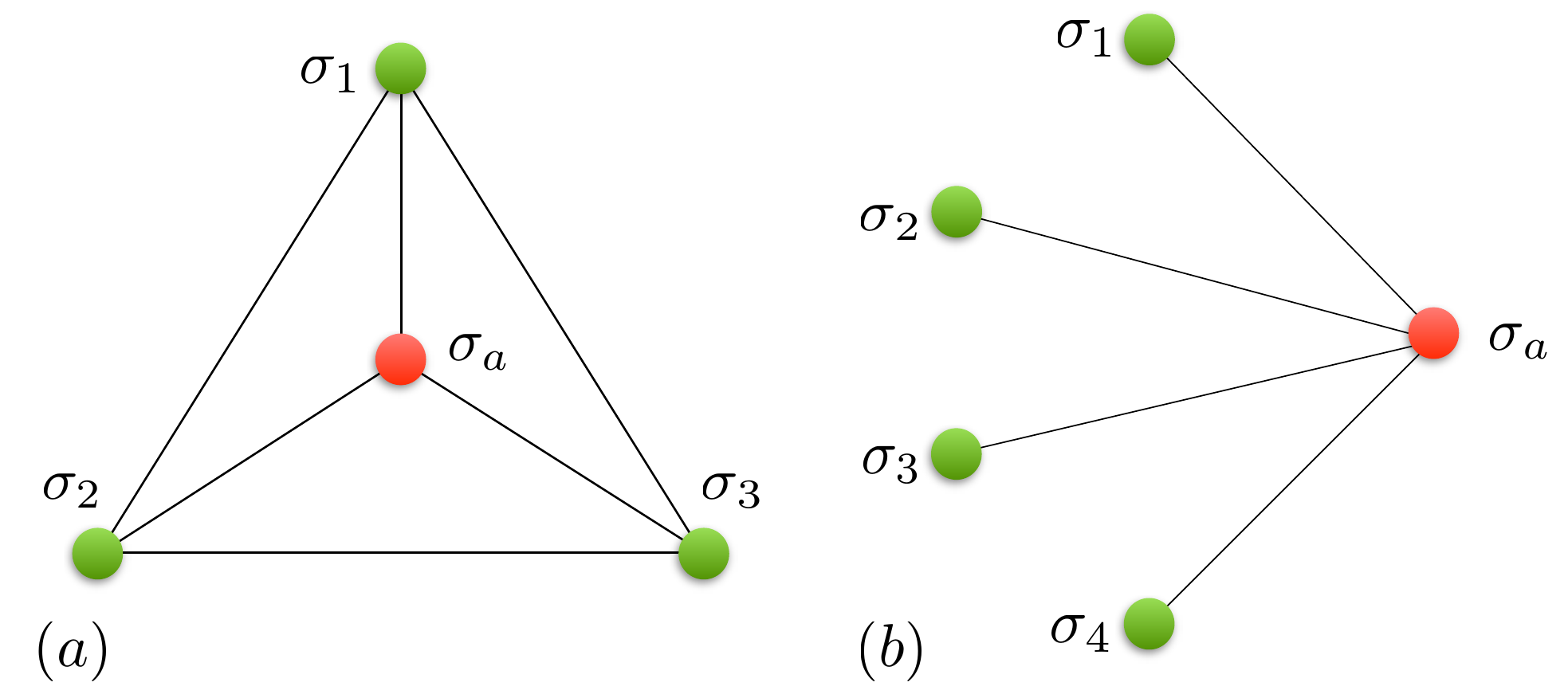}
\par\end{centering}
\caption{(a) Illustration of the abstract spin graph corresponding to the implementation of the clause $\sigma_1 \oplus \sigma_2 \oplus \sigma_3$ using only a single ancilla. (b) Illustration of the abstract spin graph for the clause $\sigma_1 \wedge \sigma_2 \wedge \sigma_3\wedge \sigma_4$. \label{fig-schematic-b} }
\end{figure}

It is worth briefly mentioning a couple of special cases, in which clauses can be expressed more efficiently than the ways discussed earlier in this section. Because the fields and couplers already act as one and two bit parity checking clauses (a field, which gives a different energy for a 1 and 0 state is nothing more than a single bit parity check), all that is needed to construct an arbitrary three bit clause is a three bit parity checking clause. Using the construction given previously for such clauses requires three ancillae. This can however be reduced to a single ancilla by choosing $h=g$, $J_a=2 J>|h|$, and $h_a=2 h$. \emph{Any} 3 bit clause can therefore be constructed using only a single ancilla. The corresponding abstract spin graph is shown in Fig.\ \ref{fig-schematic-b} (a).

Furthermore, clauses of the form $(a^{(l)}_1 \wedge ... \wedge a^{(l)}_k)$ can be expressed using only a single ancilla per sub-clause, regardless of $k$. Consider the simpler Hamiltonian of the form

 \begin{eqnarray}
\mathcal{H}_{\mathrm{simple}}^{(2)}&=&  h \sum_{i=1}^k  c(i)\,\sigma^{z}_i + 
J^a \sum_{i=1}^k c(i) \sigma^{z}_i \sigma^{z}_{a} +   h^a \sigma^{z}_{a},
\label{Hsimple}
\end{eqnarray}
in which up to the gauge choice $c(i) \in \{ -1,1\}$, $k$ logical spin variables $\sigma^{z}_i$ are coupled with equal strength $J^a$ to the same ancilla spin variable $\sigma^{z}_{a}$. The connectivity of the corresponding spin graph is shown in Fig.\ \ref{fig-schematic-b} (b) for the case $k=4$. We choose $ h^a$ so that if all of the logical bits $\sigma^{z}_i$ match $c(i)$, the ancilla bit will be down. This can be achieved as follows: Consider choosing $h_a=J^a\,k+q$. The ground state of the ancilla will be the upward orientation unless all logical bits cooperate to counteract the field in which case it will be downward.  We can further choose the couplers between the logical bits such that the energy of the ancilla-logical couplers exactly cancels the energy from the couplers between the logical qubits. We now further set $J^a=-h$ and $q = -g/2$, leading to

\begin{eqnarray}
h_a= - h\,k- g/2,  \quad J^a=-h \label{choices-and}
\end{eqnarray}
which (up to a constant offset) yields a total energy of zero iff $c(i)\sigma^z_i=+1, \forall i$ and $g$ otherwise, which is exactly the spectrum of a clause of the form  $(a^{(l)}_1 \wedge ... \wedge a^{(l)}_k)$. 

\section*{Embedding in the Chimera Graph}

\begin{figure}
\begin{centering}
\includegraphics[width=7cm]{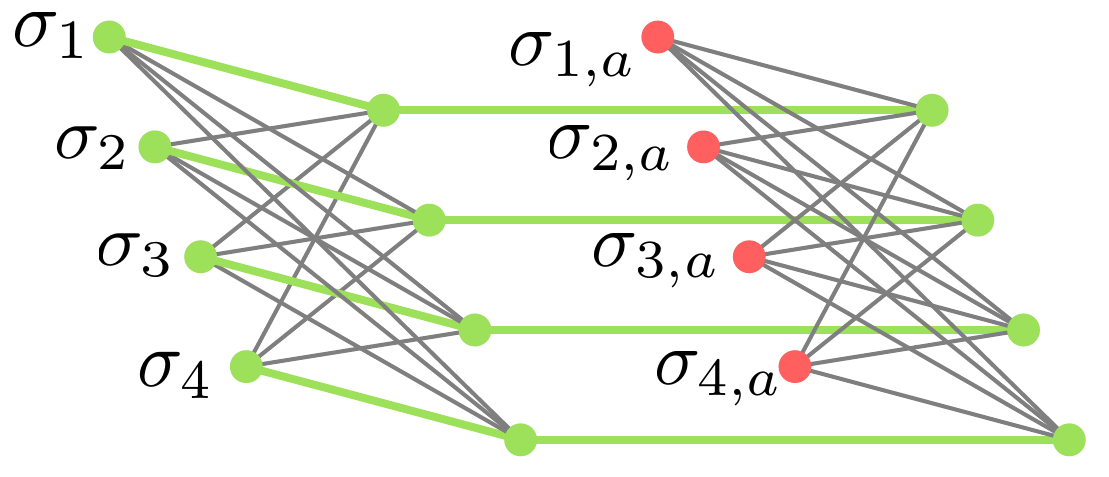}
\par\end{centering}
\caption{ Illustration of the spin graph corresponding to the minor embedding of the clause $\sigma_1 \vee \sigma_2 \vee \sigma_3 \vee \sigma_4 $ as well as the parity checking clause $\sigma_1 \oplus \sigma_2 \oplus \sigma_3 \oplus \sigma_4 $ in the Chimera graph. The unit cell on the left implements the fully connected graph amongst the for logical variables, while the unit cell on the right implements the ancillae. Strong coupling edges between vertices (shown as thick green edges) are used to ``identify'' spin variables.\label{fig-embedding_chimera} }
\end{figure}

Let us start by considering how to embed both simple clauses and parity checks represented by the connectivity graph shown in Fig.\ \ref{fig-schematic} into a patch of a Chimera graph.  This embedding is shown in Fig.\ \ref{fig-embedding_chimera}. The abstract graph shown in Fig. \ref{fig-schematic} is already reduced to have only standard two-body interactions between neighboring spins as is the case for the Chimera graph. The only difficulty in embedding this abstract graph is the fact that its connectivity is higher than that of the Chimera graph. In particular, the embedding of the clauses of order $k$ involves a fully connected graph of the $k$ logical spin variables. Higher connectivity can be achieved at the price of an overhead in the number of spin variables by `identifying' different spin variables through a very strong link. In other words, two spins are coupled through a ferro-magnetic link of strength $|J^\infty|$ which is much larger than all the other couplings, ensuring that both spins always have the same value. This can be seen in the embedding of the clauses in Fig.\ \ref{fig-embedding_chimera}, where the strong links identifying logical spin variables are shown as thick green edges.

\begin{figure}
\begin{centering}
\includegraphics[width=7cm]{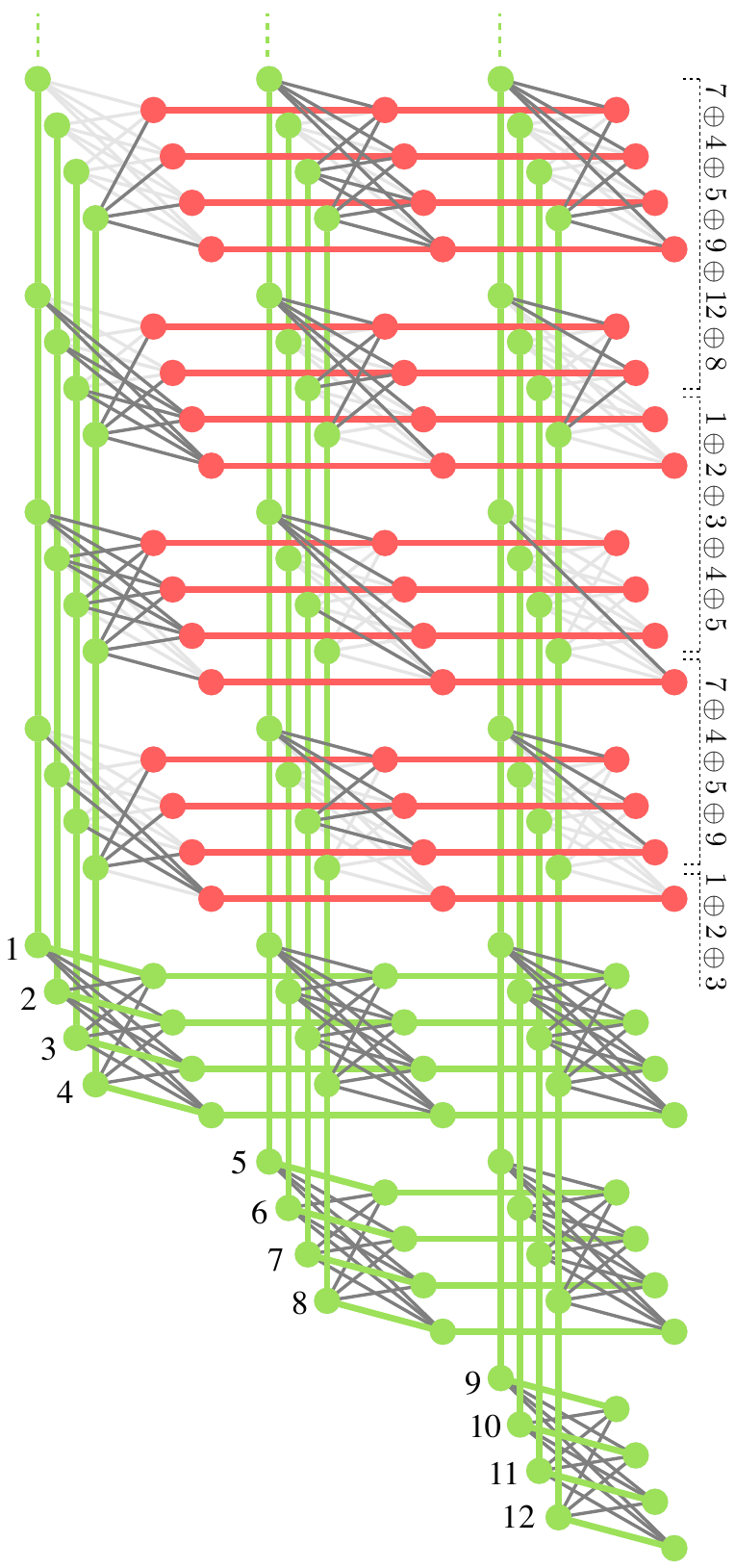}
\par\end{centering}
\caption{Embedding of a few clauses of a 12 bit problem in a Chimera graph. The actual problem is part of a turbo code involving 12 message bits and 12 parity bits (of which eight are omitted in the figure). The bottom 3 rows implement a fully connected graph amongst the 12 message variables, which allows for arbitrary 2 bit clauses.  The upper 4 rows illustrate how the ancillae can be embedded. To implement a clause using our method, the number of ancillae needs to be equal to the number of logical bits in the clause and each row of Chimera unit cells provides 4 ancillae. \label{fig2}}
\end{figure}

 To generalize the above embedding for a larger number of variables and clauses one can employ the minor embedding of a fully connected graph which was introduced in \cite{Choi2011}. Since clauses can be superimposed, it suffices to have a single minor embedding of all logical spin variables. While one could also include all ancillae in a single fully connected graph and then set unused edges to zero, this would not be very efficient. A more efficient way to do this is to extend each of the logical spin variables as a string of physical spin variables coming out of one side of the fully connected embedding of all logical spin variables, with each of the ancillae  as an embedding chain crossing all of them. This is illustrated in Fig.\ \ref{fig2}. In the lower three rows of unit cells we see a minor embedding of a fully connected graph between 12 logical spin variables. In the upper part of the figure, each of the 12 spin variables has outgoing chains of `identified' spins which are then connected to the ancillae. This plot illustrates a specific example of an embedding to perform some of the parity checks for the turbo code example given previously which is explained in detail in that section.

The method proposed in the previous paragraph works well for embedding any number of arbitrary overlapping subgraphs of the form shown in Fig.\ \ref{fig-schematic}. For problems requiring a large number of ancillae however, embeddings with this method will only occupy a long relatively thin strip of the Chimera graph. In practice, real devices tend to be designed with an aspect ratio close to 1:1, which maximizes the tree-width for given number of qubits.  Because of this, the way to embed into a real device will be to use a serpentine pattern, which uses fully connected graphs as `corners' at the end of each row and allow the embedding to efficiently fill the graph.

\section*{Scaling}

Let us now consider the scaling of the total number of physical qubits, $N_{phys}$ required to embed an instance of a max $k$-SAT problem using penalty terms of the form \eqref{eq:spectrum} only. We shall first assume that the long stretches of linear embedding chains which cross the ancillae will dominate.  In this case the total number of physical qubits will scale as, 
\begin{equation}
N_{phys}\propto N_{log}\left<k\right>  \,c
\label{eqscale_par}
\end{equation}
where $N_{log}$ is the number of logical qubits, $c$ is the number of clauses, and $\left<k\right>$ is the mean number of bits per clause. It has been demonstrated, for instance with max 2-SAT \cite{Coppersmith(2004),Santra2014} that the typical hardness of a problem is determined by the ratio $r=c/N_{log}$. If clauses are too sparse, then it will typically be easy to satisfy them all simultaneously, however if clauses are too dense, the problem again becomes easy because no solution will be able to satisfy very many of the clauses and almost any random bit string will be a good solution. Based on this reasoning, the value of $r$ which gives the hardest typical problems should not vary too much from $r \approx O(1)$ and therefore if we want scaling for the hardest problems $r$ can be treated as roughly independent of $\left<k\right>$, the mean number of bits in a clause. 

If we are interested in using annealers to solve problems drawn from a typically hard set of problems, the number of physical qubits required should scale roughly as, 

\begin{equation}
N_{phys}\propto N^2_{log}\,\left<k\right>\,r
\label{eqscale_par_hard}
\end{equation}
which, is the same scaling as the minor embedding for a fully connected graph proposed in \cite{Choi2011}.

This scaling only applies for those clauses which can be expressed as terms of the form \eqref{eq:spectrum}. In general one can imagine much more complicated clauses.  Using the more naive method of constructing a clause piecewise by penalizing each bitstring on $k$ bits can at most require $ 2^k-1$ subclauses to implement. Each of these subclauses will require an ancilla to implement. For this reason the average number of ancillae to implement naively is $\left<N_{anc}(k) \right> \leq 2^{k}-1$.  The actual value of $\left<N_{anc}(k)\right>$ is likely to be highly dependent on both the specific max-SAT problem and the algorithm to implement the clauses as discussed previously.

For arbitrary clauses, involving a maximum of $k$ bits, the worst case scenario scaling for typically hard problems is therefore,
\begin{equation}
N_{phys}\propto N^2_{log}\,r\,\left<N_{anc}(k)\right>,
\label{eqscaling_worst_hard}
\end{equation}
which scales exponentially with the clause length $k$, but still only scales quadratically with $N_{log}$, given a fixed value of $k$. This scaling still assumes we are in the typically hard regime. However, we can further calculate the absolutely worst case, which one clause needs to be constructed on every subset of $k$ or fewer bits. Assuming that $c\gg k$, then the leading order scaling in the number of ancilla required will be $N_{log}^{k}$, and the overall scaling will be, assuming $k\geq 3$,
\begin{equation}
N_{phys}\propto N_{log}^{k+1},
\label{eqscaling_worst_worst}
\end{equation}
which is still polynomial in $N_{log}$ for fixed $k$, however the rapidly growing power will place limits on practical realizations of such pathological problems. It worth pointing out here that while such problems can be mathematically constructed, it is not clear that any problems whose embedding scales like \eqref{eqscaling_worst_worst} are actually of any practical interest, or indeed even that hard problems which scale like this exist.

It is worth noting that while it is possible to implement 3 bit clauses with only a single ancilla, this technique does not have a meaningful effect on the overall scaling.

\section*{Application to Turbo Codes}  \label{sec-turbo}

As a practical example of scaling, let us consider decoding a turbo code. We restrict ourself to the basic construction and leave a detailed analysis to a forthcoming work \cite{in_prep}. While this problem is most naturally stated in terms of Ising spins, we have already shown that this problem maps to a max k-SAT. Turbo codes are a class of so-called convolutional code which have many real life applications in communications due to the ability to achieve near Shannon limit performance \cite{berrou1,berrou2,mackaybook}. 
The Hamiltonian for decoding of a general turbo code can be written in the following way,
\begin{eqnarray}
H(\rc) &=&  f \sum_{i=1}^K \rho_i \sigma_i^{z} + f \sum_{i=1}^K \rho_{i+K} \left( \mathbb{I}(\text{$i$ odd}) \prod_{j=1}^i \sigma^z_j +\right. \nonumber\\
 &&   \quad \quad  \left. +\mathbb{I}(\text{$i$ even}) \prod_{j=1}^i \sigma^z_{p(j)} \right)
\label{turbo}
\end{eqnarray}
where $p(j)$ is a random permutation, and $\rho_i\in\{-1,1\}$, with $i=1,...,2K$ the list of received values of the message and the corresponding parity checks. Furthermore, $f$ is related to the noise model; in the case of a binary symmetric channel which corrupts a value with probability $p$, one has
\begin{equation} 
f = - \frac{1}{2}\log\left(\frac{1-p}{p}\right).
\end{equation}
The values $\rho_i$, with $i=1,...,K$ represent the received values corresponding to the original message, while $\rho_i$, with $i=K+1,...,2K$ are the received values corresponding to the parity checks. The parity checks are performed in a nested structure, firstly one transmits parity checks on odd numbers of variables with respect to the original ordering of the spin variables, i.e.\ $\sigma^z_1$, $\sigma^z_1\sigma^z_2\sigma^z_3$, $\sigma^z_1\sigma^z_2\sigma^z_3\sigma^z_4\sigma^z_5$, etc. and secondly one transmits parity checks of even number of variables on the permuted variables, i.e.\ $\sigma^z_{p(1)}\sigma^z_{p(2)}$, $\sigma^z_{p(1)}\sigma^z_{p(2)}\sigma^z_{p(3)}\sigma^z_{p(4)}$, etc.

The number of ancillae required for each parity checking clause scales like the length of the clause, and there is one clause for every possible clause length in \eqref{turbo}. The number of ancilla required to decode the turbo code therefore scales as $N_{anc}\propto N_{log}^2$. The total number of physical bits will therefore scale as $N_{phys}\propto N_{log}^3$ for turbo code decoding.

To give an explicit example, Fig. \ref{fig2} shows part of the embedding of a turbo code. We have $K=12$ message variables and an equal number of parity checks. The message variables are represented by the logical spin variables (green) which are fully connected. On the outgoing chains we couple the ancillae corresponding to the parity checks. Note that parity checks of order 1 and 2 can be directly implemented on the fully connected graph. The next highest order is the parity check of order three on the variables 1, 2 and 3 which is implemented using the efficient embedding using a single ancilla. Next we have a parity check on the four variables $p(1),p(2),p(3),p(4)$, where for concreteness we chose $p(1)=7,p(2)=4,p(3)=5,p(4)=9$. The figure also shows the next two parity checks of order 5, involving variables $1,2,3,4,5$, and of order 6, involving $p(1),p(2),p(3),p(4),p(5),p(6)$, where in the example we chose $p(5)=12,p(6)=8$.

\section*{Conclusion}

We have demonstrated a method to embed max $k$-SAT (and weighted max $k$-SAT), including the problem of finding a bit-string which satisfies a maximum number of parity checks, which can be expressed in terms of clauses involving XOR. We further demonstrate a very efficient way to implement such clauses, which have important applications in communications, we discuss the specific example of turbo code decoding. 

The weighted max $k$-SAT problem is the most general satisfiability problem, so we have therefore demonstrated how \emph{any} satisfiability problem can be directly mapped into a Chimera graph. One particularly interesting application of this is parity checking, which could lead to important applications in communications.  Furthermore, the methods given here reproduce the low energy spectrum of the problem, with energies corresponding to the number of clauses which are unsatisfied. This means that as well as finding the lowest energy solution, these techniques are compatible with maximum entropy inference applications.
 
 Our method gives a direct construction of the problem Hamiltonians without the need for numerically expensive classical calculations. However the methods given here could probably be made more powerful if integrated into the already powerful and growing numerical toolset which is currently used to map problems into a Chimera for real calculations, for example in \cite{Bian2014}. The embedding illustrated here was chosen for its generality, but for specific problems it is unlikely that every ancilla will have to couple to every logical qubit, or that 2-qubit couplers between every logical qubit will be necessary. Therefore, embedding efficiency gains are likely to be possible through numerical optimization.  Furthermore it would be interesting to explore how these methods can be generalized to more sparse graphs and more general clause types.
 



\begin{acknowledgments}
N.C. was supported by  EPSRC (grant ref:  EP/L022303/1). S.Z. acknowledges support by Nokia Technologies, Lockheed Martin and the University of Oxford through the Quantum Optimisation and Machine Learning (QuOpaL) Project. P.W. was supported by Lockheed Martin and by EPSRC (grant refs: EP/K004506/1 and EP/H005544/1). SCB is supported by the EPSRC National
Quantum Technology Hub in Networked Quantum Information Technologies (grant ref:  EP/M013243/1). 
\end{acknowledgments}


\section*{Author Contributions}

SZ, NC and PW came up with the initial concept and wrote the paper, with SZ creating the figures. SB and SR helped to supervise and were involved in developing the scientific content and editing the manuscript along with all other authors.

\section*{Competing Interests}

The authors declare no competing interests.

\bibliography{}{}

\end{document}